\documentclass[10pt]{elsarticle}

\usepackage{latexsym}
\usepackage{bm}
\usepackage{upgreek}
\usepackage{mathrsfs}
\usepackage{times}
\usepackage{amsthm}
\usepackage{amssymb}
\usepackage{epsfig}
\usepackage{graphicx}
\usepackage{amsmath}
\usepackage{color}

\definecolor{purple}{rgb}{0.8,0,0.6}

\newcommand{\revision}[1]{{#1}}
\newcommand{\revisionA}[1]{{#1}}

\newcommand{\revisionG}[1]{{#1}}
\newcommand{\revisionQ}[1]{{#1}}
\newcommand{\revisionC}[1]{{#1}}

\theoremstyle{plain}
\newtheorem{theorem}{Theorem}[section]

\newtheorem{lemma}{Lemma}[section]

\theoremstyle{definition}

\theoremstyle{remark}
\newtheorem{remark}[lemma]{Remark}

\begin{document}

\title{Emergent Weyl spinors in multi - fermion systems}

\author[AALTO,ITP]{G.E.~Volovik}

\address[AALTO]{Low Temperature Laboratory, Aalto University, P.O. Box
15100,
FI-00076 Aalto, Finland}
\address[ITP]{Landau Institute for Theoretical Physics RAS, Kosygina 2,
119334 Moscow, Russia}

\author[ITEP,UWO]{M.A.~Zubkov}

\address[ITEP]{ITEP, B.Cheremushkinskaya 25, Moscow, 117259, Russia}
\address[UWO]{University of Western Ontario,  London, ON, Canada N6A 5B7}

\begin{abstract}
In Ref. \cite{Horava2005}  Ho\v{r}ava suggested, that the multi - fermion
many-body
system with topologically stable Fermi surfaces may effectively be
described (in a vicinity of the Fermi surface) by the theory with
coarse-grained fermions. The number of the components of these
coarse-grained fermions is reduced compared to the original system. Here we
consider the $3+1$ D system and concentrate on the particular case  when the
Fermi surface has co-dimension $p=3$,
i.e. it represents the Fermi point in momentum space. First we demonstrate
explicitly that  in agreement with  Ho\v{r}ava
conjecture, in the vicinity of the Fermi point
the original system is reduced to the model with two - component Weyl
spinors. Next, we generalize the construction of
 Ho\v{r}ava to the situation, when the original $3+1$ D theory contains
multi -
component Majorana spinors. In this case the system is also reduced to the
model of the two - component Weyl fermions in the vicinity of the
topologically stable Fermi point.  Those fermions experience the emergent
gauge
field and the
gravitational field given by the emergent veirbein. Both these fields (the
emergent gauge field and the emergent
gravitational field) originate from certain collective excitations of the
original system. We speculate, that the given construction may be relevant
for the high energy physics in the paradigm, in which the  Lorentz symmetry
as well as the gravitational and gauge fields are the emergent phenomena,
i.e. they appear dynamically in the low energy approximation of the
underlined high energy theory.
\end{abstract}



\maketitle

\section{Introduction}

As in particle physics, the condensed matter systems are  described by the
multi - component
fermionic fields. In addition to spin they may have Bogoliubov spin, layer
index in the multilayered $2+1$ systems, etc. In crystals the band indices
are added, and the spinor acquires infinite number of components.
In the low energy corner the effective number of degrees of freedom is
essentially reduced.
The gapped (massive) degrees of freedom are frozen out and only gapless
states survive.
The gaplessness is the fragile property, since it can be violated by
interaction between femions.
However, there exist fermionic systems, in which the  gaplessness
(masslessness) is robust to interaction.
These are the topological materials, where stability of nodes in the energy
spectrum with respect to deformations is protected by the conservation of
topological invariants of different types \cite{Horava2005}.

Examples of topologically protected zeroes in fermionic spectrum are:
Fermi surface in metals \cite{VolovikBook};
Fermi points  in $3+1$ D Weyl superfluid $^3$He-A \cite{VolovikBook} and in
$3+1$ D Weyl semimetals
\cite{Abrikosov1971,Abrikosov1998,Burkov2011,XiangangWan2011,Weylsemimetal};
Dirac points in graphene
\cite{Semenoff:1984dq,CastroNeto:2009zz};
fermionic edge modes on the surface and interfaces of the fully gapped
topological insulators
\cite{VolkovPankratov1985,HasanKane2010,Xiao-LiangQi2011} and superfuids
\cite{SalomaaVolovik1988,Volovik2009}.

The Fermi or Weyl points represent  the exceptional (conical, diabolic)
points of level crossing, which avoid the level repulsion
\cite{NeumannWigner}. Topological invariants for points at which the
branches of spectrum merge were introduced by Novikov \cite{Novikov1981}.
In our case crossing points occur in momentum space
\cite{Avron1983,Volovik1987}.

The spectrum near the point nodes typically acquires the relativistic form,
which is the consequence
of the Atyah - Bott - Shapiro construction applied to the nodes with unit
value of topological invariant
\cite{Horava2005}. This results in emergence of
effective gauge and gravitational field as collective Bose modes
\cite{Froggatt1991,VolovikBook,Volovik1986A,Volovik2011}.
 This means, that
the fermionic excitations reside in curved space - time. The geometry of
this space - time is given by the vierbein formed by
certain collective excitations of the microscopic system.

The higher values of topological invariant give rise to exotic Weyl or Dirac
fermions, with nonlinear touching
points of positive and negative energy branches. The $2+1$ D example of such
system is given by the multilayer graphene  with
the $ABC$ stacking
\cite{multilayer}. The nonlinear Dirac spectrum results in the effective
gravitational and gauge field theories,
which obey anisotropic scaling of  Ho\v{r}ava type
\cite{HoravaPRL2009,HoravaPRD2009,Horava2008,Horava2010}, see
\cite{KatsnelsonVolovik2012,KatsnelsonVolovikZubkov2013a,KatsnelsonVolovikZubkov2013b,VZ2013gr}.
The multilayer graphene also demonstrates the reduction of the degrees of
freedom at low energy.
The original tight - binding model may be described by the field theory with
the multi -
component fermionic field, which carries the spin, pseudospin, and layer
indices. Due to the specific interaction between the fermions that belong to
different layers, in the emergent low energy theory the layer
index drops out. The final effective theory operates with
the two - component spinors existing in the vicinity of each of the two
Fermi points. These spinors also carry the flavor index that corresponds
to the real spin.

The general theory that describes reduction of the fermion components and
the emergent gravity experienced by the reduced fermions is not developed so
far in sufficient details. The main progress in this direction has been made
by  Ho\v{r}ava \cite{Horava2005}, who considered
the
general case of $d+1$ dimensional condensed matter system with $d-p$
dimensional Fermi
surface ($d-p$ dimensional manifold of zeroes in the $d$ dimensional
momentum space).
The classification of the fully gapped topological materials
\cite{Schnyder2008,Kitaev2009}
can be obtained from Ho\v{r}ava classification by dimensional reduction (see
examples in \cite{VolovikBook}).

We are interested in the particular case of the $3+1$ D systems
with Fermi - point, i.e.  with the node of co-dimension $p=d=3$.
In this particular case it follows from the statement of  \cite{Horava2005},
that in the vicinity of the Fermi - point the system is effectively
described by the two - component fermion field $\Psi$. The action of this
two - component field is given by
\begin{equation}
{\cal S} = \int d \mu({\bf p}, \omega) \bar{\Psi}({\bf p}, \omega) D
\Psi({\bf p}, \omega), \label{Hamiltonian}
\end{equation}
where $\mu$ is the integration measure over momentum and frequency ${\bf p},
\omega$. It was claimed in \cite{Horava2005}, that operator $D$ contains the
construction of Atyah - Bott - Shapiro that enters the expression for the
topological invariant corresponding to the nontrivial $\pi_3(GL(n,C)) = {\bf
K}(R^3)$, where $n$ is the original number of the fermion components:
\begin{equation}
D  = e^{\mu}_a \,
\sigma^a \left(p_\mu -p^{(0)}_\mu\right)+  \ldots
\label{D}
\end{equation}
Here $p_{\mu}$ is $4$ - momentum; $e^{\mu}_a$ is an emergent vierbein;
 $p^{(0)}_\mu$ is the position of the Fermi point, whose space - time
variation
 gives rise to the effective dynamical $U(1)$ gauge field $B_{\mu}$ ;
 and dots mean
the subdominant terms, which include the emergent spin connection $C_{\mu}$.

The emergence of this equation (\ref{D}) has been advocated by Froggatt and
Nielsen
in their random dynamics theory, where the infinite number of degrees of
freedom
is reduced to $2\times 2$ subspace of Hermitian matrices (see page 147 in
the book \cite{Froggatt1991}).
In superfluid $^3$He-A this equation  (\ref{D})  has been explicitly
obtained by expansion
of the Bogoliubov-de Gennes Hamiltonian near the Weyl point
\cite{Volovik1986A};  for the expansion near Dirac point in $2+1$ D graphene
see Ref. \cite{VK2010,Manes2013,VZ2013gr}.
In both cases the complicated atomic structure of liquid and electronic structure in crystals are
reduced to
the description in terms of the effective two-component spinors, and this
supports the conjecture of Froggatt and Nielsen
and the  Ho\v{r}ava approach.

The emergence of Weyl spinor has important consequences both in the
condensed matter physics and in the high energy physics. This is because
 the Weyl fermions represent the building blocks of the Standard Model of
particle physics (SM).
 Emergence of Weyl fermions in condensed matter together with Lorentz
invariance, effective gravity and gauge fields and the topological stability
of emergent phenomena suggest  that  SM  and  Einstein theory of
gravitational field (GR) may have the status of effective theories. The
chiral elementary particles (quarks and leptons), gauge and Higgs bosons,
and the dynamical vierbein field may naturally emerge in the low-energy
corner of the quantum vacuum, provided the vacuum has topologically
protected Weyl points.

 When considering the possible emergence of SM and GR, one should resolve
between the symmetries which emerge in the low energy corner (Lorentz
invariance, gauge symmetry, etc.) and the underlying symmetry
 of the microscopic system -- the quantum vacuum.
The  discrete and continuous global symmetries of the underlying microscopic
systems influence the
topological classification producing the additional classes of system, which
are protected by the combined action of symmetry and topology
\cite{VolovikBook,Schnyder2008,Kitaev2009,Fu2011,Makhlin2013}. They also
determine the effective
symmetries emerging at low energy, such as $SU(2)$ gauge symmetry in
$^3$He-A, which follows from the
discrete $Z_2$ symmetry of the underlying high-energy theory
\cite{VolovikBook}.

Especially we are interested in the case, when
the original multi - fermion system consists of real fermions. i.e. it is
the system
of the underlying  Majorana fermions of general type not obeying Lorentz
invariance.
This case may be related both to emergent gravity and to the foundations of
quantum mechanics.
The equations of ordinary quantum mechanics  are described in terms of
complex numbers. These are  the Weyl equation;
the Dirac equation obtained after electroweak symmetry breaking, when
particles acquire Dirac masses; and finally  the Schr\"odinger equation
obtained for energies below the mass parameters. As is known, Schr\"odinger
strongly resisted to introduce
$i=\sqrt{-1}$ into his wave equations (see Yang \cite{Yang}). The imaginary
unit $i=\sqrt{-1}$ is the product of human mind,
which is mathematically convenient. However, all the physical
quantities are real, which implies that the imaginary unit
should not enter any physical equation.

This suggests that the underlying microscopic physics is described solely in
terms of the
real numbers, while the complexification occurs on the way from microscopic
to macroscopic physics,
i.e. complexification of quantum mechanics (and of the quantum field theory)
is
the emergent phenomenon that appears at low energies.
To see that we start with underlying microscopic system described in
terms of the real - valued multi - component spinor, whose evolution is
governed by the differential equation with real coefficients.
We find that if the vacuum is toplogically
nontrivial,
the low energy phenomena will be described by the emergent Weyl quantum
mechanics, which is
expressed in terms of the emergent complex numbers.

 The quantum
dynamics of the corresponding field system is described by the integral over
the $n$ - component Grassmann variables $\psi$ that does not contain
imaginary unity. In the low energy approximation the multi - component
Majorana fermions are reduced to the two - component Weyl fermions, which
descisription is given in terms of the complex - valued two component wave
function. The functional integral of $e^{ i S}$ is over the two sets of
$2$ - component Grassmann variables $\Psi$ and $\bar{\Psi}$, where $S$ is
the action for the emergent Weyl fermions $\Psi$ (and the conjugated
fermions $\bar{\Psi}$) in the presence of the emergent vierbein $e^{\mu}_a$
and emergent gauge field.

It is worth mentioning that in most of the cases the  main symmetry of the
gravitational theory (invariance under the diffeomorphisms) does not arise.
For emergence of the diffeomorphism invariance the Lorentz violation scale
must be much higher than the Planck scale. If this hierarchy of scale is not
obeyed,
in addition to Eq. (\ref{Hamiltonian}) the
effective action contains the terms that do not depend on $\Psi, \bar{\Psi}$
but depend on $e^j_k$, $C_{\mu}$ and $B_{\mu}$ directly. These terms are, in
general case, not invariant under the diffeomorphisms. That's why in the
majority of cases we may speak of the gravity only as the geometry
experienced by fermionic quasiparticles. The fluctuations of the fields
$e^j_k$, $B_{\mu}$, and $C_{\mu}$ themselves are not governed by the
diffeomorphism - invariant theory.

We shall demonstrate, that under certain reasonable assumptions the emergent
spin connection $C^{ab}_{\mu}$ in the considered systems is absent. This
means, that we deal with the emergent  teleparallel gravity, i.e. the theory
of
the varying Weitzenbock geometry
\footnote{The Riemann - Cartan space is
defined by the translational connection (the vierbein) and the Lorentz group
connection. There are two important particular cases. space is called
Riemannian if the translational curvature (torsion) vanishes. If the Lorentz
group curvature vanishes, it is called Weitzenbock space.}.

On the high - energy side the application of the given pattern may be
related to the unification of interactions in the paradigm, in which at the
extremely high energies the Lorentz - invariance as well as the general
covariance are lost. In this paradigm Lorentz symmetry, the two - component
Weyl fermions that  belong to its spinor representation, the gravitational
and gauge fields appear at low energies as certain collective excitations of
the microscopic theory.

The paper is organized as follows. In section \ref{sectHorava} we describe
the original construction of Ho\v{r}ava \cite{Horava2005} and give its proof
for
the particular case of the $3+1$ D system, in which Fermi - surface is
reduced to Fermi - point. In section \ref{sectMajorana} we generalize the
construction of section \ref{sectHorava} to the case, when the original
system contains multi - component Majorana fermions. In section
\ref{Discussion} we end with the conclusions.

\section{Emergent Weyl spinors in the system with multi - component fermions
(Ho\v{r}ava construction)}
\label{sectHorava}

\subsection{The reduction of the original multi - fermion model to the model
with minimal number of spinor components}

{Following \cite{Horava2005} we consider the condensed matter model with
$n$ - component spinors $\psi$. The partition function has the form:
\begin{equation}
Z = \int D \psi D \bar{\psi} {\rm exp}\Bigl(i\int d t \sum_{{\bf x}}
\bar{\psi}_{\bf x}(t) (i\partial_t - \hat{H}_{\rm }) \psi_{\bf x}(t)
\Bigr),\label{ZH}
\end{equation}
Here the Hamiltonian $H$ is the Hermitian matrix function of momentum
$\hat{\bf p} = - i \nabla$. We introduced here the symbol of the summation
over the points of coordinate space. This symbol is to be understood as the
integral over $d^3 x$ for continuous coordinate space. First, we consider
the particular case, when
there is no interaction between the fermions and the coefficients in the
expansion of $H$ in powers of $\bf p$ do not depend on coordinates.
We know, that there is the "repulsion" between the energy levels in ordinary
quantum mechanics. Similar situation takes place for the spectrum of $\hat
H$.
The eigenvalues of $\hat H$ are the real - valued functions of $\bf p$.}

{Several branches of spectrum for the  Hermitian operator $\hat H$ repel
each other, i.e. any small perturbation pushes apart the two crossed
branches. That's why only the minimal number of branches of its spectrum may
cross each other. This minimal number is fixed by the topology of momentum
space that is the space of parameters $\bf p$.}

{Let us consider the position ${\bf p}^{(0)}$ of the crossing of $n_{\rm
reduced}$ branches of $\hat H$. There exists the Hermitian matrix $\Omega$
such that the matrix $\tilde{H}( {\bf p}) = \Omega^+ \hat{H} \Omega$ is
diagonal. In this matrix the first $n_{\rm reduced}\times n_{\rm reduced}$
block $\hat{H}_{\rm reduced}$ corresponds to the crossed branches (i.e. all
eigenvalues of $\hat{H}_{\rm reduced}({\bf p})$ coincide at ${\bf p} = {\bf
p}^{(0)}$). The remaining block of matrix $\hat{H}_{\rm massive}$
corresponds to the "massive" branches. The functional integral can be
represented as the product of the functional integral over "massive" modes
and the integral over $n_{\rm reduced}$ reduced fermion components
\begin{equation}
\Psi_{\rm }({\bf x}) = \Pi\, \Omega \psi({\bf x}), \quad
\bar{\Psi}_{\rm }({\bf x}) = \bar{\psi}({\bf x})\Omega^+ \Pi^+
\end{equation}
Here $\Pi$ is the projector to space spanned on the first $n_{\rm reduced}$
components. }
{Let us denote the remaining components of $\psi$ by
\begin{equation}
\Theta_{\rm }({\bf x}) = (1-\Pi)\, \Omega \psi({\bf x}), \quad
\bar{\Theta}({\bf x}) = \bar{\psi}({\bf x})\Omega^+(1-\Pi^+)
\end{equation}}

{Let us denote the only eigenvalue of $\hat{H}_{\rm reduced}({\bf p}^{(0)})$
by $E_0$. The transformation
$\psi_{\bf x} \rightarrow e^{-i E_0 t} \psi_{\bf x}$ , $H({\bf p})
\rightarrow H({\bf p}) - E_0$ leaves the expression in exponent of Eq.
(\ref{ZH}) unchanged.  That's why we can always consider the matrix
${H}_{\rm reduced}$ equal to zero at the position of the branches crossing
${\bf p}^{(0)}$. We are left with the following expression for the partition
function:
\begin{equation}
Z = \int D \Psi D\bar{\Psi} D \Theta D\bar{\Theta} {\rm exp}\Bigl(i\int d t
\sum_{{\bf x}} \Bigl[\bar{\Psi}_{\bf x}(t) (i\partial_t - \hat{H}_{\rm
reduced}) \Psi_{\bf x}(t) + \bar{\Theta}_{\bf x}(t) (i\partial_t -
\hat{H}_{\rm massive}) \Theta_{\bf x}(t)\Bigr]\Bigr)\label{ZH2}
\end{equation}
where ${H}_{\rm massive} = (1-\Pi) \hat{H}  (1-\Pi^+)$.       }

{Spectrum of operator $\hat H_{\rm reduced}$ has exceptional properties
around vanishing eigenvalues. The corresponding eigenfunctions do not depend
on time.   The key point is that at low energy the integral over
$\Psi_{\rm }({\bf x})$ dominates. The other components $\Theta$ contribute
the physical quantities with the fast oscillating factors, and, therefore,
may be neglected in the description of the long - wavelength dynamics. As a
result at the low energies we may  deal with the theory that has the
following partition function:
\begin{equation}
Z = \int D \Psi D\bar{\Psi}  {\rm exp}\Bigl(i\int d t \sum_{{\bf x}}
\bar{\Psi}_{\bf x}(t) (i\partial_t - \hat{H}_{\rm reduced}) \Psi_{\bf x}(t)
\Bigr)\label{ZH3}
\end{equation}}

Here we consider the situation, when Fermi energy coincides with the value
of energy at the branches crossing. It was suggested by Froggatt and Nielsen
in their random dynamics theory, that this case may be distinguished due to
the specific decrease of particle density as follows from  the Hubble
expansion \cite{Froggatt1991}.

\subsection{Momentum space topology,  and the two -
component spinors}

\label{sectf0}

{This consideration allows to prove the Ho\v{r}ava's conjecture presented in
\cite{Horava2005}. According to this conjecture any condensed matter theory
with fermions and with the topologically protected Fermi - points may be
reduced at low energies to the theory described by the two - component Weyl
spinors. The remaining part of the proof is the consideration of momentum
space topology. It protects the zeros of $\hat H_{\rm reduced}$ (i.e. it is
robust to deformations) only when there is the corresponding nontrivial
invariant in momentum space. The minimal number of fermion components that
admits nontrivial topology is two.} This reduces the partition function to
\begin{equation}
Z = \int D \Psi D \bar{\Psi} {\rm exp}\Bigl(  i \int d t \sum_{{\bf x}}
\bar{\Psi}_{{\bf x}}(t) (i\partial_t - m^L_k(\hat{\bf p}) \hat \sigma^k  -
m(\hat{\bf p}))\Psi_{\bf x}(t) \Bigr)\label{ZH__0}
\end{equation}
where functions $m^L_k, m$ are real - valued.

Let us, in addition,  impose the CP symmetry generated
by ${\cal CP} = - i \sigma^2 $ and followed by the change ${\bf x}
\rightarrow -{\bf x}$. Its action on the spinors is:
\begin{eqnarray}
&& {\cal CP} \Psi({\bf x}) = - i \sigma^2 \bar{\Psi}^T(-{\bf x})
\end{eqnarray}
It prohibits the term with $m({\bf p})$. Thus operator $\hat H$ can be
represented as
\begin{eqnarray}
\hat H &=&   \sum_{k=1,2,3} m^{L}_k( {\bf p})\hat \sigma^k
\end{eqnarray}

The topologically nontrivial situation arises when
${ m}^{L}({\bf p})$ has the hedgehog singularity.
The hedgehog point zero is described by the topological invariant
\begin{equation}
N= \frac{e_{ijk}}{8\pi} ~
   \int_{\sigma}    dS^i
~\hat{ m}^L\cdot \left(\frac{\partial \hat{ m}^L}{\partial p_j}
\times \frac{\partial \hat{ m}^L}{\partial p_k} \right), \quad \hat{ m}^L =
\frac{{ m}^L}{|{ m}^L|}
\label{NH}
\end{equation}
where $\sigma$  is the $S^2$ surface around the point.

For the topological invariant $N=1$ in Eq.(\ref{NH}) the expansion near the
hedgehog point at $p^{(0)}_j$ in $3D$ ${\bf p}$-space gives
\begin{equation}
m^L_i({\bf p})=f_i^j(p_j-p^{(0)}_j)\,.
\label{A(K)-expansion0H}
\end{equation}
Here by $f^j_i$ we denote the coefficients in the expansion. It will be seen
below, that these constants are related to the emergent vierbein.
{As a result, Eq. (\ref{ZH__0}) has the form:
\begin{equation}
Z = \int D \Psi D\bar{\Psi} {\rm exp}\Bigl(i\int d t \sum_{{\bf x}}
\bar{\Psi}_{{\bf x}}(t) (i\partial_t - f_k^j(\hat{\bf p}_j-p^{(0)}_j) \hat
\sigma^k )\Psi_{\bf x}(t) \Bigr)\label{ZHH}
\end{equation}}

\begin{remark} \label{CPremark}
\revisionG{In the absence of the mentioned above $CP$ symmetry we have, in
addition, the function $m({\bf p})$ that is to be expanded around ${\bf
p}^{(0)}$:  $m({\bf p}) \approx f_0^j(p_j-p^{(0)}_j )$, $i,j =
1,2,3$. The new quantities $f_0^j$  are introduced here. So, in general case
we arrive at the expression for the partition function of Eq. (\ref{ZHH}),
in which the sum is over $k = 0,1,2,3$, and $j = 1,2,3$ while $\sigma^0 =
1$.  The situation here becomes much more complicated, than in the presence
of the $CP$ symmetry. Namely, when $|f^j_0 [f^{-1}]_j^a| \ge 1$, we have the
more powerful zeros of the Hamiltonian (better to say, - of its
determinant). For $|f^j_0 [f^{-1}]_j^a| > 1$ there is the conical Fermi -
surface of co-dimension $p=1$ given by the equation
\begin{equation}
f_k^j(p_j-p^{(0)}_j) = 0, \quad j = 1,2,3; \, k=0,1,2,3
\label{cfs}
\end{equation}
There exists the choice of coordinates, such that on this Fermi surface the
energy of one of the two branches of spectrum of $H$ is equal to zero. The
energy corresponding to the second branch vanishes at ${\bf p}^{(0)}$ only,
where the two branches intersect each other. However, in this situation the
first branch dominates the dynamics, and we already do not deal with the
Fermi - point scenario of the effective low energy theory. That's why the
$CP$ - invariance is important because it protects the system from the
appearance of the Fermi surface in the vicinity of the branches crossing. It
is worth mentioning, that  in the marginal case $|f^j_0 [f^{-1}]_j^a| = 1$
we deal with the line of zeros of the Hamiltonian (Fermi -
surface of co-dimension $p=2$). We do not consider here
the other marginal cases, such as that, in which ${\rm det}\, f^j_a = 0$.}
\end{remark}

\revisionG{In the following we shall imply, that there is the additional
symmetry (like the mentioned above $CP$ symmetry) that protects the system
from the appearance of the more powerful zeros in the spectrum of the
Hamiltonian (i.e. Fermi surfaces and Fermi lines). The $CP$ symmetry may be
approximate instead of exact, i.e. it may be violated by small perturbations
and the interactions. The approximate $CP$ symmetry is enough to provide the
inequality $|f^j_0 [f^{-1}]_j^a| < 1$ that restricts the appearance of the
Fermi - surfaces of co-dimension $p=1$ and $p=2$. \revisionQ{In this case we may apply
Lorentz transformation (boost) that brings the system to the reference
frame, in which $f^j_0 = 0$ for $j=1,2,3$.}  In the following the value of
${\bf p}^{(0)}$ may be interpreted as the external vector potential. The
interpretation of quantity $f^j_k$ in terms of the emergent gravitational
field will be given in the next subsection.}

\subsection{Taking into account interaction between the fermions}
\label{sectinter0}

Next, we should consider the situation, when the coefficients of expansion
of $H$ in powers of $\bf p$, depend on coordinates and fluctuate. The
original partition function for the fermions with the interaction between
them can be written as follows:
\begin{equation}
Z = \int D \psi D\bar{\psi} D\Phi {\rm exp}\Bigl(i R[\Phi] + i \int d t
\sum_{{\bf x}} \bar{\psi}_{\bf x}(t) (i\partial_t - \hat{H}(\Phi)) \psi_{\bf
x}(t) \Bigr)\label{FIn}
\end{equation}
Here the new fields that provide the interaction between the fermions are
denoted by $\Phi$. $R$ is some function of these fields.  Now operator $\hat
H$ also depends on these fields. In mean field approximation, when the
values of $\Phi$ are set to their "mean" values we come back to the
consideration of the previous subsections. However, at the end of the
consideration the fluctuations of the fields $\Phi$ are to be taken into
account via the fluctuations of the field $f^a_k$ and the Fermi - point position ${\bf p}^{(0)}$.

{Let us consider  for the simplicity the low energy effective theory with
only one emergent Weyl fermion. The interaction between the particles
appears when the fluctuations of $p^{(0)}_k$ and $f^j_k$ are taken into
account. {We assume, that these fluctuations are long - wave, so that the
corresponding variables should be considered as if they would not depend on
coordinates.} Nevertheless, in the presence of the varied field $\Phi$ the
time reversal symmetry is broken.  As a result the partition function of the
theory receives the form
\begin{equation}
Z = \int D \Psi D \bar{\Psi} D\Phi {\rm exp}\Bigl(i R[\Phi]\Bigr) {\rm
exp}\Bigl(  i \int d t \sum_{{\bf x}} \bar{\Psi}_{{\bf x}}(t) (i\partial_t -
m^L_{\Phi, k}(\hat{\bf p}) \hat \sigma^k  - m_{\Phi}(\hat{\bf p}))\Psi_{\bf
x}(t) \Bigr)\label{ZH__0}
\end{equation}
Here
\begin{equation}
m^{L}_{\Phi, i}({\bf p})\approx e \, e_i^j(p_j - B_j ),  \quad
m_{\Phi}({\bf p}) \approx B_0 + e\, e_0^j(p_j-B_j ),\quad i,j = 1,2,3
\label{A(K)-expansion0120}
\end{equation}
The appearance of the field $B_0$ reflects, that in the presence of
interaction the value of energy at the position of the crossing ot several
branches of spectrum may differ from zero. We represented the quantity
$f^j_i$  of Eq. (\ref{A(K)-expansion0H}) (that
depends now on the coordinates) as $f_i^j = e \, e_i^j$, where the
fluctuating long - wave fields $e[\Phi],B[\Phi]$ depend on the primary
fields $\Phi$. This representation for $f_i^j$ is chosen in this way in
order to interpret the field $e_i^j$ as the vierbein. We require
$e^0_a = 0$ for $a=1,2,3$, and $e \times e_0^0=1$. Here $e^{-1} = e_0^0
\times {\rm det}_{3\times 3}\, e^i_a = e_0^0$ is equal to the determinant of
the vierbein $e^i_a$. In the mean field approximation, $\Phi$ is set to its
mean value $\Phi_0$, while $B_0[\Phi_0] = 0$, and $e\,
e^k_a[\Phi_0] = f^k_a$, where  variable $f$ was introduced in sect.
\ref{sectf0}.   \revisionG{It is implied (see remark \ref{CPremark}), that
the approximate CP symmetry is present, that may be slightly violated by the
interactions. This means, that the values of $e_0^j$ are suppressed compared
to the values of $e^j_k$ for $k = 1,2,3$. This allows to keep the Fermi
point in the presence of interactions.}

{As a result, the partition function of the model may be rewritten as:
\begin{equation}
Z = \int D \Psi D\bar{\Psi} D e^i_k D B_k e^{i S[e^j_a, B_j,
\bar{\Psi},\Psi]}\label{ZI}
\end{equation}
with
\begin{eqnarray}
S &=&   S_0[e, B] + \frac{1}{2} \Bigl(\int d t \,e\,  \sum_{{\bf x}}
\bar{\Psi}_{{\bf x}}(t)  e_a^j  \hat \sigma^a  \hat D_j  \Psi_{\bf x}(t) +
(h.c.)\Bigr),\label{Se0}
\end{eqnarray}
where the sum is over $a,j = 0,1,2,3$ while $\sigma^0 \equiv 1$, and $\hat
D$ is the covariant derivative that includes the $U(1)$ gauge field $B$.
$S_0[e, B]$ is the part of the effective action that depends on $e$ and $B$
only.}

\begin{remark}\label{Boostremark}
\revisionQ{It is worth mentioning, that to write the expressions for the
functional integral Eq. (\ref{ZI}) and the
expression for the action Eq. (\ref{Se0}) is not enough to define the field
system. Besides, we are to impose boundary conditions on the fields.
Typically, the anti - periodic in time boundary conditions are imposed on
the spinor fields in quantum field theory. These boundary  conditions
correspond to the choice of vacuum, in which all states with negative energy
are occupied. This is important to point out the reference frame, in which
these anti - periodic boundary conditions in time are applied. Here and
below we always imply, that these boundary conditions are imposed in the
synchronous reference frame, i.e. in the one, in which the mean values
$\langle e^j_0\rangle$ vanish for $j = 1,2,3$. }
\end{remark}

\begin{remark}
Eq. (\ref{Se0}) is reduced to Eq. (\ref{ZH__0}) with $m,m^L$ given by Eq.
(\ref{A(K)-expansion0120}) if the particular gauge (of the emergent
$SO(3,1)$) is fixed. In this gauge $e^0_j = 0$ for $j = 1,2,3$. Besides, we
rescale time in such a way, that $e\, e^0_0=1$.  This means, that the term
$S_0$ contains the corresponding gauge fixing term. Even modulo this gauge
fixing the theory given by Eq. (\ref{Se0}) is not diffeomorphism -
invariant. The fermionic term alone would become diffeomorphism - invariant
if the spin connection of zero curvature is added. Then, in addition Eq.
(\ref{Se0}) is  to be understood as the result of the gauge fixing
corresponding to vanishing spin connection.  \revision{In some cases $S_0[e,
B]$ may be neglected, and only the second
term of Eq. (\ref{Se0}) contributes the dynamics. Then the fields
$e^{\mu}_k$ and
$B_{\mu}$ may be identified with the true gravitational field (vierbein) and
the true gauge field correspondingly (modulo mentioned above gauge fixing).
Their effective action is obtained as
a result of the integration over the fermions. It is worth mentioning, that
in most of the known condensed matter systems with Fermi - points (say, in
$^3$He-A) we cannot neglect the term $S_0[e, B]$. That's why the given
opportunity in the condensed matter theory remains hypothetical.}

\end{remark}

{Recall, that we have considered the long - wavelength fluctuations of the
emergent fields $B$ and $e$. That is we neglected the derivatives of these
fields. In the fermion part of the action in Eq. (\ref{Se0}) there are no
dimensional parameters.
The only modification of this action that is analytical in $B, e$ and their
derivatives and that does not contain the dimensional parameters is if the
covariant derivative $D$ receives the contribution proportional to the
derivative of $e$. That's why, even for the  non - homogenious variations of
$e$ and $B$ in low energy approximation we are left with effective action of
the form of Eq. (\ref{Se0}) { if the value of the emergent electromagnetic
field is much larger than the order of magnitude of quantity $|\nabla
e^k_a|$}. Such a situation takes place, for example for the consideration of
the emergent gravity in graphene \cite{VZ2013gr}.}

Let us formalize the consideration of the given section as the following
theorem.
\begin{theorem}\label{theoremH}
The multi-fermion system without interaction between the particles in the
vicinity of the Fermi - point (Fermi surface of co-dimension $p=3$) is reduced to the model that is described by
the two - component Weyl fermions described by partition function Eq.
(\ref{ZH__0}). \revisionG{In addition, we require, that the (approximate) CP
symmetry is present. This symmetry
prohibits the appearance of the Fermi surfaces of co-dimension $p=1$ and $p=2$ and results in the suppression
of the values of $m({\bf p})$ compared to the
values of $m^L({\bf p})$. }
The nontrivial momentum space topology with the topological invariant of Eq.
(\ref{NH}) equal to unity provides that the effective low energy theory has
the partition function of Eq. (\ref{ZHH}) with some constants $f^j_k$ that
depend on the underlying microscopic theory.

When the interaction between the original fermions in this system is taken
into account (while momentum space topology remains the same as in the non -
interacting theory), the partition function of the low energy effective
theory receives the form of Eq. (\ref{ZI}) with the effective action Eq.
(\ref{Se0}). This is the partition function of Weyl fermion in the presence
of the emergent vierbein $e$ and the emergent $U(1)$ gauge field $B$. Both
these fields represent certain collective excitations of the microscopic
theory. (It is assumed, that the value of the emergent electromagnetic field
is much larger than the order of magnitude of quantity $|\nabla e^k_a|$.)
\end{theorem}

\begin{remark}
One can see, that in the considered long wave approximation the emergent
spin connection $C_{\mu}$ does not arise. That's why we deal with the
emergent teleparallel gravity described by the veirbein $e^i_j$ only.
\end{remark}

The given theorem represents the main statement given without proof in
\cite{Horava2005} in a more detailed and elaborated form (for the particular
case of $3+1$ D system with Fermi - surface reduced to the Fermi - point).
We considered only one Fermi point. This case also corresponds to the
situation, when there exist several Fermi points, but the corresponding
collective excitations do not correlate with each other. The situation, when
the correlation is present is more involved. We make a remark on it at the
end of section \ref{sectMajorana}.

\section{Emergent Weyl spinors in the system of multi - component Majorana
fermions}
\label{sectMajorana}

In this section we consider the generalization of the problem considered in
the previous section to the case, when the original system contains multi -
component Majorana fermions.

\subsection{Path integral for Majorana fermions}
\label{SectFuncInt}

On the language of functional integral the evolution in time of the field
system is given by the correlations of various combinations of the given
fields. The lagrangian density for $n$ - component Majorana fermions $\psi$
can be written in the form:
\begin{equation}
L_{\rm Majorana} =   \psi_{\bf x}^T(t) (i\partial_t + i\hat{A}) \psi_{\bf
x}(t),
\end{equation}
\revisionC{where $\hat A$ is the arbitrary operator that may be highly non -
local.}  First, we consider the situation, when there is no interaction
between the original Majorana fermions. This means, that operator $\hat A$
does not depend on the other fields.  As a
result the partition function is represented as
\begin{equation}
Z = \int D \psi {\rm exp}\Bigl( -\int d t \sum_{{\bf x}} \psi_{\bf x}^T(t)
(\partial_t + \hat{A}) \psi_{\bf x}(t) \Bigr)\label{FI}
\end{equation}
Various correlators of the field $\psi$ are given by
\begin{equation}
\langle \psi_{{\bf x}_1}(t_1)  \psi_{{\bf x}_2}(t_2)  ...  \psi_{{\bf
x}_2}(t_2) \rangle= \int D \psi {\rm exp}\Bigl(- \int d t \sum_{{\bf x}}
\psi_{\bf x}^T(t)  (\partial_t + \hat{A}) \psi_{\bf x}(t) \Bigr) \psi_{{\bf
x}_1}(t_1)  \psi_{{\bf x}_2}(t_2)  ...  \psi_{{\bf
x}_2}(t_2)\label{correlator}
\end{equation}
Here $\psi$ is the $n$ - component anti - commuting variable. The Majorana
nature of the fermions is reflected by the absence of the conjugated set of
variables $\bar{\psi}$ and the absence of the imaginary unit in the
exponent. The dynamics of the system is completely described by various
correlators of the type of Eq. (\ref{correlator}). It is worth mentioning
that the complex numbers do not enter the dynamics described by Eq.
(\ref{correlator}). It can be easily seen, that if $A$ is linear in the
spacial derivatives, and is represented by the product $ B_a \nabla^a$,
\revisionC{where $B_a$ do not depend on coordinates}, then $B_a$ should be
symmetric. (For the anti - symmetric $B_a$ expression
$\sum_{\bf x}\psi_{\bf x}^T  B_a \nabla^a \psi_{\bf x}$ vanishes.) We feel
this instructive to give the representation of the partition function of
Eq.(\ref{FI}) in terms of the analogues of the energy levels.

{We consider the functional integral over real fermions basing on the
analogy with the integral over complex fermions (see
\cite{PhysRevD.12.2443}).
We start from the partition function of Eq. (\ref{FI}).
In lattice discretization the differential operator $\hat A$ is represented
as the skew - symmetric  $Nn\times Nn$ matrix, where $N$ is the total number
of the lattice points while $n$ is the number of the components of the
spinor $\psi$. As a result there exists the orthogonal $Nn\times Nn$
transformation $\Omega$ that brings matrix $\hat A$ to the block - diagonal
form with the $2\times 2$ blocks of the form
\begin{equation}
E_k \hat \beta = E_k \left(\begin{array}{cc}0 & -1 \\1 & 0
\end{array}\right)
\end{equation}
with some real values $E_k$.
We represent $\psi$ as $\psi(x,t) = \sum_n
c_{ a, n}(t) \Psi_{a,n}(x)$, where $a = 1,2$, and  ${\hat A}$ has the
above block - diagonal form in the basis of $\Psi_{a,n}$. These vectors are
normalized to unity ($\int\!\!d^3x\,{\Psi}_{an}^T \Psi_{an} = 1$). Further,
we represent
\begin{equation}
Z= \int\!\! \,d c\,
{\rm exp}
\Big(-
\sum\limits\sb{\eta,n}T\,
{c}_{-\eta,n}^T [-i\eta+ E_n{\hat \beta}] c_{\eta,n}
\Big),
\end{equation}
where the system is considered
with the anti-periodic in time boundary conditions:
$\psi(t+T, x) =-\psi(t, x)$.
We use the decomposition
\begin{equation}\label{dec-0}
c_n(t) = \sum_{\eta=\frac{\pi}{T}(2k+1),\, k \in Z} e^{-i \eta t}
c_{\eta,n}.
\end{equation}
Integrating out the Grassmann variables $c_n$ we come to:
\begin{eqnarray}
\label{G_p_h_E}
&&
Z= \prod_{\eta>0}
\prod_{n}
\bigl((\eta+E^{}_n)(-\eta+E^{}_n)T^2\bigr)=\prod_{\eta}
\prod_{n}
\bigl((\eta+E^{}_n)T\bigr)
 =\prod\sb{n}\cos\frac{ T E^{\varphi}_n}{2},
\end{eqnarray}}

{The values $E_n$ depend on the parameters of the Hamiltonian,
with the index $n$ enumerating these values.
Eq. (\ref{G_p_h_E}) is derived as follows.
Recall that
in \eqref{dec-0}
the summation is over
$\eta=\frac{\pi}{T}(2k+1)$.
The product over $k$ can be calculated as in \cite{PhysRevD.12.2443}:
\begin{equation}
\prod_{k\in Z}
\Big(1 + \frac{ E_n T}{\pi(2k+1)}\Big)
=
\cos\frac{E^{}_n T}{2},
\label{DET0}
\end{equation}
Formally the partition function may be rewritten as
\begin{equation}
Z=  {\rm Det}^{1/2} \Bigl[ \partial_t + \hat A \Bigr]= \prod_n \cos\frac{E_n
T}{2}
\end{equation}
The explanation that the square root of the determinant appears is that
operator $\Bigl[ \partial_t + \hat A \Bigr]$ itself being discrcetized
becomes the skew - symmetric matrix. Via the orthogonal transformations it
may be made block - diagonal with the elementary $2\times 2$ blocks. In the
latter form the functional integral is obviously equal to the square root of
the determinant because for the 2 - component spinor $\eta$
\begin{equation}
\int d \eta \, {\rm exp}\Bigl[\eta^T \left(\begin{array}{cc}0 & -a \\a & 0
\end{array}\right) \eta\Bigr] = a = {\rm Det}^{1/2}\left(\begin{array}{cc}0
& -a \\a & 0 \end{array}\right)
\end{equation}
We get (see also \cite{PhysRevD.12.2443}):
\begin{eqnarray}
&&
\hskip -40pt
Z
=
\sum_{\{K_n\} = 0,1}
{\rm exp}
\Bigl(
 \frac{i T}{2} \sum_n
E_n-i T \sum_n K_n E_n
\Bigr)
 \label{G_p_h_E_0}
\end{eqnarray}}

Following \cite{PhysRevD.12.2443},
we interpret Eq. (\ref{G_p_h_E_0}) as follows. $K_n$
represents the number of occupied states with the energy $E_n$. These
numbers may be $0$ or $1$. The term $\sum_n E_n$ vanishes
if values $E_n$ come in pairs with the opposite signs (this occurs when the
time reversal symmetry takes place).
We can rewrite the last expression in the form, when the integer numbers
represent the numbers of occupied states of positive energy and the holes in
the sea of occupied negative energy states:
\begin{eqnarray}
Z(T) = \sum_{\{K_n\} = 0,1} {\rm exp}\Bigl(  \frac{i T}{2} \sum_n |E_n|-i T
\sum_n K_n |E_n| \Bigr)
 \label{G_p_h_E_}
\end{eqnarray}

After the Wick rotation we arrive at
\begin{eqnarray}
Z(-i/{\cal T}) = \sum_{\{K_n\} = 0,1} {\rm exp}\Bigl(  \frac{1}{2 {\cal T}}
\sum_n |E_n|-\frac{1}{{\cal T}} \sum_n K_n |E_n| \Bigr),
 \label{G_p_h_E_2}
\end{eqnarray}
where $\cal T$ is temperature. This shows, that in equilibrium the
configuration dominates with the vanishing numbers $K_n$. This corresponds
to the situation, when all states with negative energy are occupied. This
form of vacuum is intimately related with the anti - periodic in time
boundary conditions imposed on $\psi$. The other boundary conditions would
lead to the other prescription for the occupied states in vacuum.

%
%
\revisionC{The values $E_n$ are given by
the solution of the system of equations
\begin{equation}
\begin{array}{c}
\hat A \zeta_1 = E \zeta_2\\
\hat A \zeta_2 = -E \zeta_1
\end{array}
\end{equation}
for the pair $\zeta^1, \zeta^2$ of the real - valued $n$ - component wave
functions. Alternatively, we
may solve equation
\begin{eqnarray}
0 &=& [\hat A + \partial_t] {\xi}\label{sch}
\end{eqnarray}
Here the the complex - valued
 $n$ - component wave function ${\xi} = \zeta_1 + i\zeta_2$ has the
particular dependence on time
${\xi}(x,t) = \tilde{\xi}(x)e^{- i E_n t} $. However, Eq. (\ref{sch}) does
not contain imaginary unity. Therefore, we may consider its real - valued
solutions. These solutions may be interpreted as the time - dependent real -
valued spinor wave functions of Majorana fermions. It is worth mentioning,
that there are no such real valued wave functions that would correspond to
definite energy.}

\subsection{Repulsion of fermion branches $\rightarrow $ the reduced number
of fermion species at low energy.}
\label{TopologyZeroes}

{The notion of  energy  in the theory described by operator $\hat A$ may be
based on the definition of the values $E_n$ given above. Besides, we may
introduce the notion of energy scale $\cal E$  as the typical factor in the
dependence of various dimensionless physical quantities $q$ on time:  $q
\approx f({\cal E} t)$, where $f$ is a certain dimensionless function of
dimensionless argument such that its derivatives are of the order of unity.
With this definition of energy it can be shown, that at low energies only
the minimal number of fermion components effectively contributes the
dynamics. Below we make this statement explicit and present the sketch of
its proof. }

{As it is explained in Sect. \ref{SectFuncInt}, operator $\hat{A}$  in
lattice discretization is given by the skew - symmetric  $Nn\times Nn$
matrix, where $N$ is the total number of the lattice points while $n$ is the
number of the components of the spinor $\psi$. As a result there exists the
orthogonal $Nn\times Nn$ lattice transformation $\hat \Omega$ that brings
matrix $\hat A$ to the block - diagonal form with the $2\times 2$ blocks of
the form $E_k \hat \beta = E_k \left(\begin{array}{cc}0 & -1 \\1 & 0
\end{array}\right)$ with some real values $E_k$. In the continuum language
matrix $\Omega$ becomes the operator that acts as a $n\times n$ matrix,
whose components are the operators acting on the coordinates. There are
several branches of the  values of $E_k$. Each branch is parametrized by the
$3D$ continuum parameters.
Several branches of spectrum of $E_k$ repel each other because they are
the eigenvalues of the Hermitian operator. This repulsion means, that any
small perturbation pushes apart the two crossed branches. That's why only
the minimal number of branches of its spectrum may cross each other. This
minimal number is fixed by the topology of momentum space  (see below, sect.
\ref{sectf}).}

{As it was mentioned, there exists the orthogonal operator $\hat \Omega$ (it
conserves the norm $\int d^3 x \chi^T_x \chi_x$) such that the operator
\begin{equation}
{A}^{{\rm block}\,{\rm diagonal}} = \hat \Omega^T A \hat \Omega\label{omega}
\end{equation}
 is given by the block - diagonal matrix with the elementary $2\times 2$
blocks:
\begin{equation}
{A}^{{\rm block}\,{\rm diagonal}} = \left(\begin{array}{cccc}\hat \beta
E_1({\cal P}) & 0 & ...& 0
\\0 & \hat \beta E_2({\cal P}) &  ...& 0 \\
... & ... & ... & ...\\
0 & ... & 0 & \hat \beta E_n({\cal P})
\end{array} \right)\label{blockdiag}
\end{equation}
Here we denote by $\cal P$ the three - dimensional vector that parametrizes
the branches of spectrum and the basis vector functions that correspond to
the given form of $\hat A$. The first $n_{\rm reduced}$ values $E_k$
coincide at ${\cal P} = {\bf p}^{(0)}$. This value is denoted by $E_0 =
E_1({\bf p}^{(0)}) = E_2({\bf p}^{(0)})=...$. The first $2n_{\rm
reduced}\times 2n_{\rm reduced}$ block ${A}^{{\rm block}\,{\rm
diagonal}}_{\rm reduced}$ corresponds
to the crossed branches.
The remaining block of matrix ${A}^{{\rm block}\,{\rm diagonal}}_{\rm
massive}$ corresponds to the
"massive" branches. The functional integral can be represented as the
product of the functional integral over "massive" modes and the integral
over $2 n_{\rm reduced}$ reduced fermion components
\begin{equation}
\Psi_{\rm }({\cal P},t) = e^{E_0 \, \hat \beta \, t } \Pi\,  \psi({\cal
P},t)
\end{equation}
Here by $\hat \beta$ we denote $2n_{\rm reduced}\times 2n_{\rm reduced}$ matrix
$\beta \otimes 1$, while $\Pi$ is the projector to space spanned on the
first $2n_{\rm reduced}$ components.
Let us denote the remaining components of $\psi$ by
\begin{equation}
\Theta_{\rm }({\cal P},t) = (1-\Pi)\, \Omega \psi({\cal P},t)
\end{equation}
 We  arrive at
\begin{eqnarray}
Z &=& \int D \Psi D \Theta {\rm exp}\Bigl(- \int d t \sum_{{\cal P}}\Bigl[
{\Psi}_{\cal P}^T(t) e^{ E_0 \, \hat \beta \, t }(\partial_t + \hat{A}^{{\rm
block}\,{\rm diagonal}}_{\rm
reduced}({\cal P}))e^{ -E_0 \, \hat \beta \, t } \Psi_{\cal P}(t)
\nonumber\\&&
+  {\Theta}_{\cal P}^T (\partial_t + \hat{A}^{{\rm block}\,{\rm
diagonal}}_{\rm massive}({\cal
P}) )\Theta_{\cal P} \Bigr]\Bigr),\label{FI3}
\end{eqnarray}
where ${A}^{{\rm block}\,{\rm diagonal}}_{\rm reduced}({\cal P}) =  \Pi
{A}_{}({\cal P})\Pi^T$ while ${A}^{{\rm block}\,{\rm diagonal}}_{\rm
massive}({\cal P}) =  (1-\Pi) {A}_{}({\cal P})(1-\Pi^T)$. }

{The exponent in Eq. (\ref{FI3}) contains the following term that
corresponds to the contribution of the fermion fields defined in a  vicinity
of ${\cal P} = {\bf p}^{(0)}$:
\begin{equation}
{\cal A}_{p^{(0)}} = \int d t \sum_{{\cal P}, k = 1...n_{\rm reduced}}
{\Psi}_{k,{\cal P}}^T(t) (\partial_t +  \beta [E_k ({\cal P}) - E_k({\bf
p}^{(0)})]) \Psi_{k,{\cal P} }(t), \label{AH}
\end{equation}
We have the analogue of the $2n_{\rm reduced}\times 2n_{\rm reduced}$
hamiltonian $H({\cal P}) = [E_k ({\cal P}) - E_k({\bf p}^{(0)})]$ that
vanishes at ${\cal P} = {\bf p}^{(0)}$. Following Sect. \ref{SectFuncInt}
we come to the conclusion, that in the expression for the partition function
Eq. (\ref{G_p_h_E_}) the small values of energies  dominate (when the
negative energy states are occupied), and these energies correspond to the
reduced fermions $\Psi$. It is important, that in order to deal with vacuum,
in which negative energy states for Eq. (\ref{AH}) are occupied we need to
impose the antiperiodic boundary conditions in time on  $\Psi$ (not on the
original fermion field $\psi$).
The other components $\Theta$ contribute the physical quantities with the
fast oscillating factors because they are "massive", i.e. do not give rise
to the values of $E_n$ from the vicinity of zero. Therefore, these degrees
of freedom may be neglected in the description of the long - wavelength
dynamics.

\revision{Any basis of the wave functions is related via an orthogonal
operator $\tilde{\Omega}$ to the basis of the wave functions, in which
$\hat{A}_{\rm reduced}$ has the form of the block - diagonal matrix (Eq.
(\ref{blockdiag})). We require, that $\tilde{\Omega}$ commutes with
$\hat \beta$ for the transformation to the basis associated with the
observed low energy coordinates. This observed coordinate space may differ
from the primary one, so that  $\tilde{\Omega}$ is not equal to
$\hat{{\Omega}}$ of Eq. (\ref{omega}). This new coordinate space in not the
primary notion, but the secondary one. $[\tilde{\Omega},\beta]=0$ is
the requirement, imposed on the representation of the theory, that allows to
recover the usual Weyl spinors and the conventional quantum mechanics with
complex - valued wave functions (see the next subsection). We denote the new
coordinates by $\bf Z$ to distinguish them from the original coordinates
$\bf x$, in which the partition function of Eq. (\ref{FI}) is written. In
this new basis $\hat A_{\rm reduced}$ is given by the differential operator.
It is expressed as a series in powers of derivatives with real - valued
$2n_{\rm reduced}\times 2n_{\rm reduced}$ matrices as coefficients. From
$[\tilde{\Omega},\beta]=0$ it follows, that in this basis $[\hat
A_{\rm reduced} , \hat \beta] = 0$.}

\subsection{The reduced $4$ - component spinors}

\subsubsection{Analytical dependence of $A_{\rm reduced}$ on $\cal P$}

{In Section \ref{TopologyZeroes} it was argued that the number of fermion
components at low energies should be even. The minimal even number that
admits nontrivial momentum space topology (see below) is $4$.
That's why we consider the effective low energy four - component spinors.
\revisionC{This corresponds to the crossing of the two branches of the
energy.}

\revisionC{The $2$ values $E_k$
coincide at ${\cal P} = {\bf p}^{(0)}$. The corresponding value of $E_{1,2}$
is denoted by ${ E_0} =
E_1({\cal P}^{(0)}) = E_2({\cal P}^{(0)})$. The first $4 \times 4$ block
${A}^{{\rm block}\,{\rm diagonal}}_{\rm reduced}$ of Eq. (\ref{blockdiag})
corresponds
to the crossed branches.
The remaining block of matrix ${A}^{{\rm block}\,{\rm diagonal}}_{\rm
massive}$ corresponds to the
"massive" branches. The Fermi point appears at ${\bf p}^{(0)}$ if chemical
potential is equal
to $E_0$. Then the four reduced components dominate the functional integral
while the remaining "massive" components decouple and do not influence the
dynamics. The form ${A}^{{\rm block}\,{\rm diagonal}}_{\rm reduced} = {\rm
diag}\Bigl( E_1({\cal P})\hat{\beta} , E_2({\cal P})\hat{\beta}\Bigr)$ of the reduced matrix is exceptional. It is related by the
$4\times 4$ orthogonal transformation  $\Omega^{\prime}$ that commutes with
$\hat{\beta}$ with the $4\times 4$ matrix $A_{\rm reduced}({\cal P})$ of a
more general form. In this form $A_{\rm reduced}({\cal P})$ also commutes
with $\hat{\beta}$. In general case the dependence of $A_{\rm reduced}({\cal
P})$ on $\cal P$ is analytical. This is typical for the functions that are
encountered in physics. The non - analytical functions represent the set of
vanishing measure in space of functions. However, this is not so for the
exceptional block - diagonal form ${A}^{{\rm block}\,{\rm diagonal}}_{\rm
reduced}$ in case of non - trivial topology that protects the levels
crossing.}

\subsubsection*{Example}

\revisionC{Let us illustrate this by the example, in which
\begin{equation}
A_{\rm reduced}({\cal P}) = \hat{\beta} {\cal P}_a \Sigma^a  \label{ared}
\end{equation}
Here  the three real - valued $4\times 4$ $\Sigma$ - matrices form the basis
of the
$su(2)$ algebra and have the representation in terms of the three complex
Pauli matrices:
\begin{eqnarray}
\Sigma^1= \sigma^1 \otimes 1~,~  \Sigma^2= i_{\rm eff} \Sigma^1 \Sigma^3~,~
\Sigma^3=\sigma^3 \otimes 1\,
\label{sigma_matrices}
\label{gamma-sigma}
\end{eqnarray}
There exists the orthogonal matrix $\Omega^{\prime}$ that brings $A$ to the
block - diagonal form:
\begin{equation}
A^{{\rm block}\,{\rm diagonal}}_{\rm reduced}({\cal P}) =\sigma^3 \otimes i
\tau^2 \, \sqrt{\sum_a{\cal P}_a{\cal P}_a}
\end{equation}
One can see, that in the form of Eq. (\ref{ared}) the matrix $A_{\rm
reduced}$ is analytical at ${\cal P} = 0$ while in the block - diagonal
representation it is not.}

In the following, speaking of the low energy dynamics, we shall always
imply, that $\hat A_{\rm reduced}$ is discussed, and shall omit the
superscript "$\rm reduced $". We shall refer to space of parameters $\cal P$
as to generalized momentum space.
The zeros of $\hat A$ in this space should be topologically protected; i.e.
they must be
robust to deformations.

\subsubsection{Introduction of new coordinate space }

\revisionC{Let us identify the quantities $\cal P$ with the eigenvalues of
operator $\hat {\cal P} = - \hat{\beta}  \, \frac{\partial}{\partial
{\bf Z}}$. Here by $\bf Z$ we denote the new coordinates. They do not
coincide with the original coordinates $\bf x$. This means, that the fields
local in coordinates $\bf x$ are not local in coordinates $\bf Z$ and vice
versa.}

\subsubsection*{$1+1$ D example}

\revisionC{We illustrate the appearance of the new coordinates $\bf Z$ by
the following simple example. Let us consider the two - component Majorana
spinors in $1+1$ dimensions with original non - local operator $\hat A$
given by
\begin{equation}
\hat A = {\rm exp} \Bigl( - \hat{G} \alpha \Bigr)
\left(\begin{array}{cc}\partial_{\bf x} & 0 \\0 &  \partial_{\bf x}
\end{array}\right)\,{\rm exp} \Bigl(  \hat{G} \alpha \Bigr),
\end{equation}
where $\alpha$ is parameter while the integral operator $\hat G$ is given by
\begin{equation}
[\hat G \phi]({\bf x})= \int d {\bf y} \,f({\bf x} - {\bf y})\,  \hat
\sigma^1 \, \phi({\bf y})
\end{equation}
with some odd function $f({\bf x})$.
This operator is well - defined for the functions $\phi$ that tend to zero
at infinity sufficiently fast.}

\revisionC{Our aim is to find the two representations:}

\revisionC{1)Generalized momentum space, where $\hat{A} = E({\cal P})\,
\hat{\beta}$ for a certain function $E({\cal P})$ of generalized momenta
$\cal P$.}

\revisionC{2)New space with coordinates $\bf Z$, related to momentum space
via identification ${\cal P} = - \hat{\beta} \partial_{\bf Z}$.}

\revisionC{This aim is achieved via the following operator
\begin{equation}
\hat \Omega = {\rm exp} \Bigl( - \hat{G} \alpha \Bigr)
\end{equation}
It is orthogonal and brings $\hat A$ to the form corresponding to the new
coordinates $\bf Z$:
\begin{equation}
\hat \Omega^T \hat A \hat \Omega= \left(\begin{array}{cc}  \partial_{\bf
Z }&0 \\ 0 &  \partial_{\bf Z}
\end{array}\right) = \hat{\beta} \hat{\cal P}
\end{equation}
This defines the new coordinates $\bf Z$, in which operator $\hat A$ is
proportional to $\hat{\beta}$. Space of coordinates $\bf Z$ differs from
space of coordinates $\bf x$ just like conventional momentum space differs
from the conventional coordinate space: the functions local in one space are
not local in another one and vice versa. In generalized momentum space
operator $\hat A$ receives the form $\hat{A} = E({\cal P}) \hat{\beta}$ with
$E({\cal P}) = {\cal P}$.}

\subsubsection{How the fermion number conservation reduces the general form
of $\hat A$ for $3+1$ D Majorana fermions}

\revisionC{It was argued, that for the low energy effective fermion fields
in new coordinate space operator $\hat A$ has the form of the series in
powers of the derivatives with the $4\times 4$ real valued constant matrices
as coefficients.  Moreover,   the reduced operator $\hat A$
commutes with $\beta = \left(\begin{array}{cc}0 & -1\\1 & 0\end{array}
\right)$.}
\revisionC{The latter condition may be identified with the fermion number
conservation, that
is rather restrictive. Below we describe the general form of the $4\times 4$
operator $\hat A$ that may be expanded in powers of derivatives with real -
valued constant matrices as coefficients. It may always be considered as
skew - symmetric ($\sum_{\bf x}\chi_1^T \hat A \chi_2=-\sum_{\bf x}\chi_2^T
\hat A \chi_1$ for real - valued spinors $\chi_{1,2}$, i.e. $\hat A^T = -
\hat A$) because the combination $\sum_{\bf x}\psi^T \hat B \psi$ vanishes
for any symmetric operator $\hat B$ and Grassmann valued fields $\psi$. We
shall demonstrate how the fermion number conservation reduces the general
form of such skew - symmetric operator.} Let us introduce the two commuting
momentum operators:
\begin{equation}
\hat {\cal P}_{\beta} = - \hat{\beta} \, \nabla, \quad \hat {\cal
P}_{\alpha} = - \hat{\alpha} \, \nabla
\end{equation}
where
\begin{eqnarray}
&&  \hat \beta = -1\otimes \hat \tau_3 \hat \tau_1 = -1\otimes i \tau_2,~~
\hat \alpha = -\hat \sigma_3 \sigma_1 \otimes 1 = - i \sigma_2 \otimes 1
\end{eqnarray}}

{The two commuting operators $\hat {\cal P}_{\beta}$ and $\hat {\cal
P}_{\alpha}$ have common real - valued eigenvectors corresponding to their
real - valued eigenvalues.}
{Matrix $A$ can be represented as the analytical function
\begin{eqnarray}
\hat A &=&  {\cal F} (\hat {\cal P}_{\beta}, \hat {\cal P}_{\alpha}, \hat
L^k, \hat S^k ),
\end{eqnarray}
where
\begin{eqnarray}
&&  \hat L^k=(\hat \sigma_1 \otimes \hat \beta, -\hat \alpha \otimes 1,
\hat \sigma_3 \otimes \hat \beta),\nonumber \\
&&  \hat S^k=(\hat \alpha \otimes \hat \tau_1, -\hat 1\otimes \beta,  \hat
\alpha \otimes \hat \tau_3 ),\nonumber \\
\label{10matrices}
\end{eqnarray}
More specifically, it can be represented as
\begin{eqnarray}
\hat A &=&   \sum_{k=1,2,3} m^{L}_k( {\cal P}_{\beta})\hat L^k  +
\sum_{k=1,2,3}m^{S}_k( {\cal P}_{\alpha})\hat S^k \nonumber\\ && +
m^I_1({\cal P}_{\beta}) \hat I^1 - m^I_2({\cal P}_{\alpha}) \hat I^1 +
m^I_3({\cal P}_{\beta}) \hat I^3 - m^I_4({\cal P}_{\alpha}) \hat I^3  +
m^o({\cal P}_{\beta}) \hat \beta\,,
\label{Ageneral}
\end{eqnarray}
Here
\begin{eqnarray}
&& \hat I^1 = \hat \sigma_1 \otimes \hat \tau_3, \quad \hat I^3 = \hat
\sigma_3 \otimes \tau_3
\end{eqnarray}
while  $m^{L}_k({\cal P}), m^{S}_k({\cal P}),m^I_k({\cal P}), m^o({\cal P})$
are real - valued  functions of the momenta ${\cal P}$. Functions
$m^{I}_k({\cal P}), m^{o}_k({\cal P})$ are odd;  $\hat L^k$ and $ \hat S^k$
are the generators of the two $SO(3)$ groups; $\hat \beta$ and $\hat \alpha$
are real antisymmetric matrices that commute with all $\hat L^k$ (or $\hat
S^k$) correspondingly; $\hat I^k$ are the matrices that commute  with $\hat
\alpha \otimes \hat \beta$ but do not commute with either of $\hat \alpha$
and $\hat \beta$.
(Notice, that $\hat \beta \hat L^2 =  \hat \alpha \hat S^2 = -\alpha \otimes
\beta$. That's why odd part of the function $m^{S}_2$ may be set equal to
zero. )}

\revision{According to our condition operator $\hat A$ commutes with matrix
$ \hat \beta = 1 \otimes (-i \tau_2)$. The
coordinates of new emergent coordinate space are denoted by $\bf Z$.} Matrix
$\hat \beta$ anticommutes with $
\hat{S}_k$, $k=1,3$ and $\hat{I}_k$, $k=1,2,3,4$. {Yet another way to look
at this symmetry is to require, that the momentum defined as
$\hat {\cal P}_{\beta} = - \hat \beta \frac{\partial}{\partial {\bf Z}}$ is
conserved, i.e. commutes with $\hat A$.} This requirement reduces the
partition function to
\begin{equation}
Z = \int D \Psi {\rm exp}\Bigl(  -\int d t \sum_{{\bf Z}} {\Psi}^T_{{\bf
Z}}(t) (\partial_t + i_{\rm eff} m^L_k(\hat {\cal P}_{\beta}) \hat \Sigma^k
+ i_{\rm eff} m(\hat {\cal P}_{\beta}))\Psi_{\bf Z}(t) \Bigr)\label{Z__0}
\end{equation}
\revisionC{where $m({\cal P}_{\beta}) = m^o({\cal P}_{\beta}) - m^S_2({\cal
P}_{\alpha})$.} We introduced the effective $4\times 4$ imaginary unit
\begin{equation}
i_{\rm eff}=\hat \beta\,,
\label{effective_i}
\end{equation}
Thus operator $\hat A$ can be represented as the analytical function of
${\cal P}_{\beta}$ and $\hat L^k$ only:
$
\hat A =  {\cal F} (\hat {\cal P}_{\beta},  \hat L^k)
$.
Here we have introduced (see Eq. (\ref{sigma_matrices})) the $4\times 4$
matrices forming the quaternion
units $\Sigma_k$, that can be
represented in terms of the $2\times 2$ complex Pauli matrices.
Matrix $1\otimes \tau_3$ becomes the operator of complex conjugation.

\subsubsection{CP - symmetry and topology of zeroes}
\label{sectf}

First, let us impose the CP symmetry generated by ${\cal CP} = - i
\sigma^2 \tau^3 = \hat S^3$ and followed by the change ${\bf Z}
\rightarrow -{\bf Z}$. Its action on the spinors is:
\begin{eqnarray}
&& {\cal CP} \psi({\bf Z}) = - i \sigma^2 \tau^3 \psi(-{\bf Z})
\end{eqnarray}
It prohibits the term with $m({\cal P})$. Thus operator $\hat A$ can be
represented as
\begin{eqnarray}
\hat A &=&  {\cal F} (\hat {\cal P}_{\beta},  \hat L^k) =  \sum_{k=1,2,3}
m^{L}_k( {\cal P}_{\beta})\hat L^k
\end{eqnarray}

The topologically nontrivial situation arises when
${ m}^{L}({\cal P})$ has the hedgehog singularity.
The hedgehog point zero is described by the topological invariant
\begin{equation}
N= \frac{e_{ijk}}{8\pi} ~
   \int_{\sigma}    dS^i
~\hat{ m}^L\cdot \left(\frac{\partial \hat{ m}^L}{\partial p_j}
\times \frac{\partial \hat{ m}^L}{\partial p_k} \right),
\quad \hat{ m}^L = \frac{{ m}^L}{|{ m}^L|}\label{N}
\end{equation}
where $\sigma$  is the $S^2$ surface around the point.
For the topological invariant $N=1$ in Eq.(\ref{N}) the expansion near the
hedgehog point at $P^{(0)}_j$ in $3D$ ${\cal P}$-space gives
\begin{equation}
m^L_i({\cal P})=f_i^j({\cal P}_j-P^{(0)}_j)\,.
\label{A(K)-expansion0}
\end{equation}

{As a result, Eq. (\ref{Z}) has the form:
\begin{equation}
Z = \int D \Psi {\rm exp}\Bigl(-  \int d t \sum_{{\bf Z}} {\Psi}^T_{{\bf
Z}}(t) (\partial_t + i_{\rm eff} f_k^j(\hat{\cal P}_j-P^{(0)}_j) \hat
\Sigma^k )\Psi_{\bf Z}(t) \Bigr)\label{Z}
\end{equation}}
{Operator $\hat A$ can be written in the basis of the eigenvectors of $\cal
P$. As a result, we arrive at the $4\times 4$ matrix function of real
variable $\cal P$:
$A({\cal P})= i_{\rm eff} m^L_k( {\cal P}) \hat \Sigma^k$.
The matrix $A({\cal P})$ near the node has the form
\begin{equation}
A({\cal P})=  i_{\rm eff} \Sigma^i f_i^j({\cal P}_j-{\cal P}^{(0)}_j)\,.
\label{A(K)sigma}
\end{equation}}
In the presence of the CP symmetry the topological invariant responsible for
the singularity can be written
analytically if one considers the extended matrices $A(P_{\mu})\equiv
A({\cal
P},{P_4})=P_4+ A({\cal P})$. As a result, for generator of $\pi_3$ we have
(compare with the generator of $\pi_3(R_n)$ for $n>3$ on page 133 of Ref.
\cite{Whitehead1942}):
\begin{equation}
N = \frac{e_{\alpha\beta\mu\nu}}{48\pi^2}~
{\bf Tr}\int_\sigma   dS^\alpha
~ A^{-1} \partial_{p_\beta} A~ A^{-1}  \partial_{p_\mu}  A ~A^{-1}
\partial_{p_\nu}  A\,.
\label{TopInvariantA}
\end{equation}
Here $\sigma$ is the $S^3$ spherical surface around the node in 4D
$p_\mu$-space.

As in section \ref{sectf0} in the absence of CP symmetry we should introduce
the new variables $f_0^j$ and imply summation over $k = 0,1,2,3$ in Eq.
(\ref{Z}), where $\Sigma^0$ is identified with unity matrix. In this case
Eq. (\ref{TopInvariantA}) does not represent the topological invariant.
\revisionG{According to remark \ref{CPremark} we require, that the $CP$
symmetry may be violated only slightly. The explicit meaning of the word
"slightly" is given in remark \ref{CPremark}. This provides, that the more
powerful manyfold of zeros - the Fermi surface - does not appear in the
vicinity of the Fermi point.}

\revisionC{\begin{remark}
Up to the CPT transformation (that is the overall inversion
$t\rightarrow -t$, ${\bf Z} \rightarrow - {\bf Z}$) the mentioned above CP
symmetry coincides with the time inversion  transformation $\cal T$. (This
will become clear below, when we represent these emergent four - component
Majorana spinors in the form of the two - component left - handed Weyl spinors.) For the
original multi - component Majorana fermions the CPT transformation
understood as $t\rightarrow -t$, ${\bf Z} \rightarrow - {\bf Z}$ may already
not be the symmetry.
The time reversal transformation T for the original multi - component
fermions may be defined as the composition of $t\rightarrow -t$ and a
certain transformation of the multi - component spinor $\psi \rightarrow
{\cal T} \psi$, such that ${\cal T}^2 = -1$ and its action on the reduced
$4$ - component fermions is given by ${\cal T} = {\cal CP} = - i\sigma^2
\tau^3$.  The CP transformation of the original multi - component spinors
may be defined as ${\rm CPT} \times {\rm T}$. The CP transformation of the
low energy emergent fermions may originate, for example,  from  CP or T
symmetry of the original multi - fermion system.
\end{remark}}

\subsubsection{Propagator, Hamiltonian and Schr\"odinger equation}
\label{PropagatorHamiltonian}

It follows from the functional integral representation, that one can
introduce the propagator (the Green's function) and the Hamiltonian. In the
presence of the CP - symmetry (when $f_0^j = 0$) we have:
\begin{equation}
G^{-1}= i_{\rm eff} A(P_\mu)\equiv - H_{\cal P} +  i_{\rm eff}P_4\,.
\label{GreenFunction}
\end{equation}
This means that the Green's function here is determined on the imaginary
axis, i.e. it is the Euclidean Green function.
In terms of the Green's function the topological invariant $N$ in
Eq.(\ref{TopInvariantA}) has the following form:
\begin{equation}
N = \frac{e_{\alpha\beta\mu\nu}}{48\pi^2}~
{\bf Tr}\int_\sigma   dS^\alpha
~ G\partial_{p_\beta} G^{-1}
G\partial_{p_\mu} G^{-1} G\partial_{p_\nu}  G^{-1}\,,
\label{MasslessTopInvariant3D}
\end{equation}
where $\sigma$ is $S^3$ surface around the Fermi point in
$(P_1,P_2,P_3,P_4)$ space.

If the Hamiltonian belongs to topological class $N=1$ or $N=-1$, it
can be adiabatically deformed to the Weyl Hamiltonian for the right-handed
and left-handed fermions respectively:
 \begin{equation}
H=N\left(\Sigma^1 P_x + \Sigma^2 P_y + \Sigma^3 P_z\right) ~~, ~~N= \pm 1
\,,
\label{Weyl}
\end{equation}
where the emergent Pauli matrices $\Sigma^i$ describe the emergent
relativistic spin.

The matrix $i_{\rm
eff}$,  which commutes with the Hamiltonian, corresponds to the imaginary
unit
in the time dependent Schr\"odinger equation.
The latter is obtained, when $p_4$  is substituted by the operator of time
translation,
$p_4\rightarrow \partial_t$:
\begin{equation}
i_{\rm eff} \partial_t \chi=  H\chi\,.
\label{Schroedinger}
\end{equation}

{The whole wave dynamics may be formulated in terms of real functions only.
The Hamiltonian is expressed through the momentum operator $\hat{\cal
P}_{\beta}$. Its eigenvalues are parametrized by the eigenvalues $\cal P$ of
momentum, the projection $n = \pm 1$ of emergent spin $\hat \Sigma$ on
vector $m({\cal P})$, and the eigenvalue $C = \pm 1$ of the conjugation
operator  $\hat{C} = 1\otimes \tau_3$:
\begin{equation}
\mid  C, n,  {\cal P}\rangle \equiv \Bigl[e^{\frac{1}{2} i_{\rm eff}
\hat{\Sigma} \phi[m({\cal P})]}\times e^{i_{\rm eff} {\cal P} {\bf x}}\Bigr]
\mid C \rangle \otimes \mid n \rangle,
\end{equation}
where  $\mid  C \rangle = \frac{1}{2}\left(\begin{array}{c} 1 + C\\ -1 + C
\end{array}\right)$, and $\mid  n \rangle =
\frac{1}{2}\left(\begin{array}{c} 1 + n\\ -1 + n  \end{array}\right)$,
while  rotation around the vector $\phi$ by the angle equal to its absolute
value transforms a unit vector directed along the third axis into the one
directed along $m({\cal P})$. Vectors $\mid  C, n,  {\cal P}\rangle$ are the
eigenvectors of Hamiltonian correspondent to the eigenvalues $E = C |m({\cal
P})|$. Once at $t = 0$ the wave function is given by $\mid  C, n,  {\cal
P}\rangle$, its dependence on time is given by:
\begin{equation}
\chi(t) = e^{-i_{\rm eff} C |m({\cal P})|t} \mid  C, n,  {\cal P}\rangle
\end{equation}}

\subsection{Interaction between the fermions. }
\label{sectthird}
\subsubsection{Effective action for reduced fermions}

In this subsection we take into account the interactions between the
original Majorana fermions. We consider  for the simplicity the low energy
effective theory
with only one emergent Weyl fermion. The consideration is in general similar
to that of section
\ref{sectinter0}. \revisionC{However, there is the important complication
related to the Majorana nature of the original fermions.} The partition
function for the fermions with the
interaction between them can be written in the form:
\begin{equation}
Z = \int D \psi D\Phi {\rm exp}\Bigl(-R[\Phi] - \int d t \sum_{{\bf x}}
\psi_{\bf x}^T(t) (\partial_t + \hat{A}(\Phi)) \psi_{\bf x}(t)
\Bigr)\label{FIn}
\end{equation}
Again, the new fields that provide the interaction between the fermions are
denoted by $\Phi$. $R$ is some function of these fields. \revisionA{The
fields $\Phi$ are assumed to be bosonic. All existing fermionic fields of
the system are included into $\psi$. For the applications in condensed
matter physics the function $R$ is allowed to be complex  - valued. However,
the situation may be considered, when $R$ is real - valued function. In this
situation the functional integral Eq. (\ref{FIn}) does not contain imaginary
unity at all, which means, that the corresponding dynamics may be naturally
described without using complex numbers.}  Matrix $\hat A$
also depends on $\Phi$. \revisionC{When the values of $\Phi$ are set to
their
"mean" values $\Phi = {\Phi}_0$ we come back to the consideration of the
system without
interaction. In this system the reduced fields $\Psi$ and massive fields
$\Theta$ are defined.}

\revisionC{The next step is to take into account the fluctuations of the
fields
$\Phi$. We write again the effective action in terms of the fields $\Theta$
and $\Psi$. However, now the cross terms appear in the action that
correspond to the transition between the two. Besides, the operator $A_{\rm
reduced}$ depends on the fields $\Phi$ and does not necessarily commute with
$\hat{\beta}$. Integrating out $\Theta$ we arrive at the effective action
for the reduced four - component fields $\Psi$ and $\Phi$. This effective
action in general case contains the products of more, than two components of
$\Psi$, but those combinations are suppressed at low energies because the
fields $\Theta$ are massive.}

\revisionC{As a result we come to the partition function
\begin{eqnarray}
Z &=& \int D \Psi D\Phi {\rm exp}\Bigl( - \int d t \sum_{{\bf Z}}
{\Psi}^T_{\bf Z}(t) (\partial_t + A_{\rm reduced}[{\Phi}] )\Psi_{\bf Z}(t)
\Bigr)
\label{Z__0}
\end{eqnarray}
Now operator $\hat A[\Phi]$ does not necessarily commute with $\beta$. As a
result $\hat A[\Phi]$ has the general form of Eq. (\ref{Ageneral}) with
functions $m$ that depend on $\Phi$. We assume, that these fluctuations are
long - wave, so that the functions $m$ should be considered as if they would
not depend
on coordinates. }
Besides, we define the new two component spinors starting from the four -
component spinor $
\Psi = \Bigl(\Psi^1,  \Psi^2,  \Psi^3,  \Psi^4 \Bigr)^T$.
Those two - component spinors are given by
\begin{equation}
\Upsilon({\bf x}) = \left(\begin{array}{c}\Psi^1({\bf x}) + i \Psi^2({\bf
x}) \\
\Psi^3({\bf x}) + i \Psi^4({\bf x}) \end{array} \right), \quad
\bar{\Upsilon}({\bf x}) = \left(\begin{array}{c}\Psi^1({\bf x}) - i
\Psi^2({\bf
x}) \\ \Psi^3({\bf x}) - i \Psi^4({\bf x}) \end{array}
\right)^T\label{barPsi}
\end{equation}

\revisionC{In terms of these new spinors the partition function receives the
form:
\begin{equation}
Z = \int D \Upsilon D\bar{\Upsilon} D \Phi  e^{-R[\Phi] + i S[\Phi,
\bar{\Upsilon},\Upsilon]}\label{ZPSI}
\end{equation}
with
\begin{eqnarray}
S &=&   \frac{1}{2} \Bigl(\int d t \,  \sum_{{\bf Z}}
\bar{\Upsilon}_{{\bf Z}}(t)  (i \partial_t -  m^L_{\Phi, k}(\hat {\bf p})
\hat \sigma^k  - m_{\Phi}(\hat {\bf p})) \Upsilon_{\bf Z}(t) +
(h.c.)\Bigr) + S_{\Upsilon \Upsilon},\label{Se20}
\end{eqnarray}
where $ m^L_{\Phi, k}$ and $ m_{\Phi}$ are some real functions of momenta
${\bf p} = - i \nabla$ while the term $S_{\Upsilon \Upsilon}$ contains
various combinations of $\Upsilon^A\Upsilon^B$ and $\bar{\Upsilon}_C
\bar{\Upsilon}_D$:
\begin{eqnarray}
S_{\Upsilon \Upsilon} &=&   -\frac{1}{2} \Bigl(\int d t \,  \sum_{{\bf Z}}
{\Upsilon}_{{\bf Z}}(t)  ( {u}^L_{\Phi, k}(\hat {\bf p})
\hat \sigma^k  +  {u}_{\Phi}(\hat {\bf p})) \Upsilon_{\bf Z}(t)
+(h.c.)\Bigr)
\label{Se202}
\end{eqnarray}
with some complex - valued functions $u^L_k,u$. Recall, that without the
fluctuations of $\Phi$ (when we set $\Phi$ equal to its average $\Phi_0$)
the term $S_{\Upsilon \Upsilon}$ does not appear. }

\subsubsection{Fermion number conservation. Emergent gauge field and
emergent vierbein.}
{\revisionC{As in section \ref{sectinter0} we expand functions $m^{L}_{\Phi,
i}$ and $m_{\Phi, i}$ around the Fermi - point and take into account that
the parameters of the expansion fluctuate:
\begin{equation}
m^{L}_{\Phi, i}({\cal P})\approx e \, e_i^j({\cal P}_j  - B_j ),  \quad
m_{\Phi}({\cal P}) \approx e\, B_0 + e\, e_0^j({\cal P}_j - B_j ),\quad i,j
=
1,2,3
\label{A(K)-expansion012}
\end{equation}
We represented here $f_i^j = e \, e_i^j$. The
fluctuating long - wave fields $e[\Phi],B[\Phi]$ depend on the primary
fields $\Phi$. This representation for $f_i^j$ is chosen in this way in
order to interpret the field $e_i^j$ as the vierbein. This means, that we
require
\begin{equation}
e^0_a = 0, \, {\rm  for} \, a=1,2,3; \quad e \times e_0^0=1; \quad e^{-1} =
e_0^0 \times {\rm det}_{3\times 3}\, e^i_a = e_0^0 =  {\rm det}_{4\times 4}
e^i_a
\end{equation}
If $\Phi$ is set to its mean value $\Phi_0$, we need
\begin{equation}
B_0[\Phi_0] =  0,\quad e\, e^k_a[\Phi_0] = f^k_a,
\end{equation}
 where  variable $f$ was introduced in sect. \ref{sectf}.}

 \revisionC{Besides, we expand complex - valued functions $u^L_k,u$ around
the Fermi point:
 \begin{equation}
u^{L}_{\Phi, i}({\cal P})\approx e\, W_i+ e \, q_i^j({\cal P}_j  - B_j ),
\quad
u_{\Phi}({\cal P}) \approx e\, W_0 + e\, q_0^j({\cal P}_j - B_j ),\quad i,j
=
1,2,3
\label{A(K)-expansion0122}
\end{equation}
with complex - valued $W_i, q_i^j$. We set $q^0_a = 0, \, {\rm  for} \,
a=0,1,2,3$.
As a result the partition function of
the low energy effective theory receives the form
\begin{equation}
Z = \int D \Upsilon D\bar{\Upsilon} D e D B D q D W  e^{i S[e,B,q,W] + i
S[e,B,q,W,
\bar{\Upsilon},\Upsilon]}\label{ZPSI00}
\end{equation}
with the action  given by
\begin{eqnarray}
S &=&   \frac{1}{2} \int d t \,e\,  \sum_{{\bf Z}}\Bigl(
\bar{\Upsilon}_{{\bf Z}}(t)  e_a^j  \hat \sigma^a i \hat D_j  \Upsilon_{\bf
Z}(t) + i \epsilon_{AB}{\Upsilon}^A_{{\bf Z}}(t)    \Upsilon^B_{\bf Z}(t) W_2
\nonumber\\&&+ {\Upsilon}_{{\bf Z}}(t)  q_1^j  \hat \sigma^1 i \hat D_j
\Upsilon_{\bf Z}(t)+{\Upsilon}_{{\bf Z}}(t)  q_3^j  \hat \sigma^3 i \hat D_j
\Upsilon_{\bf Z}(t)+ {\Upsilon}_{{\bf Z}}(t)  q_0^j   i \hat D_j
\Upsilon_{\bf Z}(t)
+(h.c.)\Bigr) ,\label{Se00}
\end{eqnarray}
where the sum is over $a,j = 0,1,2,3$ while $\sigma^0 \equiv 1$, and $\hat
D$ is the covariant derivative that includes the $U(1)$ gauge field $B$.
$S_0[e, B,q,W]$ is the part of the effective action that depends on bosonic
fields
only. The second term of Eq. (\ref{Se00}) contains the combination of Weyl
spinors entering the Majorana mass terms. The other fermion number breaking
terms do not have the interpretation within the model of Weyl spinors in the
presence of the gravitational field.   }

\revisionC{This interpretation does appears when we
imply, that there exists the mechanism that suppresses those fluctuations of
the fields $\Phi$ of Eq. (\ref{FIn}) that break the fermion number
conservation for the reduced Weyl fermions $\Upsilon, \bar{\Upsilon}$ (i.e.
forbids the appearance of terms proportional to $\Upsilon_A {\Upsilon}_B$
and $\bar{\Upsilon}^A \bar{\Upsilon}^B$). This formulation of the fermion
number conservation is equivalent to the requirement, that the operator
$\hat A$ acting on the four - component spinors $\Psi$ commutes with $\hat
\beta$.}
Then, similar, to section \ref{sectinter0} we come to the following theorem:
\begin{theorem}
\label{theoremA}
The system of multi - component Majorana fermions without interaction
between the particles in the vicinity of the Fermi - point is reduced to the
model that is described by the two - component Weyl fermions with partition
function Eq. (\ref{Z__0}). \revisionG{In addition, we require, that the
(approximate) CP symmetry is present. This symmetry
prohibits the appearance of the Fermi surface and results in the suppression
of the values of $m({\cal P})$ compared to the
values of $m^L({\cal P})$. }
The nontrivial momentum space topology with the topological invariant of Eq.
(\ref{N}) equal to unity provides that the effective low energy
theory has the partition function of Eq. (\ref{Z}) with some quantities
$f^j_i$ that depend on the underlined microscopic theory.

\revisionC{We assume, that in the presence of interactions the fermion
number of the coarse - grained fermions remains conserved while momentum
space topology is the same as in
the non - interacting theory. The fluctuations of the original bosonic field
$\Phi$ are supposed to be long - wave.}
Then {\it there exists new coordinate space} (we
denote the new coordinates by $\bf Z$), in which the partition function of
the low energy effective theory receives the form
\begin{equation}
Z = \int D \Upsilon D\bar{\Upsilon} D e D B  e^{i S[e,B] + i S[e,B,
\bar{\Upsilon},\Upsilon]}\label{ZPSI}
\end{equation}
with the action given by
\begin{eqnarray}
S &=&   \frac{1}{2} \int d t \,e\,  \sum_{{\bf Z}}\Bigl(
\bar{\Upsilon}_{{\bf Z}}(t)  e_a^j  \hat \sigma^a i \hat D_j  \Upsilon_{\bf
Z}(t) +
(h.c.)\Bigr) ,\label{Se}
\end{eqnarray}
$S_0[e, B]$ is the part of the effective action that depends on the fields
$e$ and $B$
only.
Both these fields represent certain collective excitations of the
microscopic theory. (It is assumed, that the value of the emergent
electromagnetic field is much larger than the order of magnitude of quantity
$|\nabla e^k_a|$.) \revisionQ{As well as in the previous section (remark
\ref{Boostremark}) we impose the antiperiodic boundary conditions in time on
the spinor fields in the synchronous reference frame, where $\langle
e^j_0\rangle = 0$ for $j=1,2,3$. }
\end{theorem}

\begin{remark}
\revisionA{If the functional $R[\Phi]$ of Eq. (\ref{FIn}) is real - valued,
then the appearance of the term $iS[e,B]$ in exponent of Eq. (\ref{ZPSI})
(with real - valued $S[e,B]$) requires some comments. Let us explain how
this may occur in principle by the consideration of the following example.
We start from Eq. (\ref{ZPSI}), and rewrite it as:
\begin{equation}
Z = \int D e^i_k D B_k e^{i S_0[e, B]} {\cal Z}[e^i_k, B_k],\label{ZPSI11}
\end{equation}
where
\begin{equation}
{\cal Z}[e^i_k, B_k] = \int D \Upsilon D\bar{\Upsilon} e^{ i S[e^j_a, B_j,
\bar{\Psi},\Psi]}\label{ZPSI01}
\end{equation}
If there exists the transformation of fields $ e^j_a, B_j,
\bar{\Upsilon},\Upsilon$ such that $\cal Z$ remains invariant while $S_0$
changes the sign, then we have:
\begin{equation}
Z = \int D e^i_k D B_k {\rm cos}(S_0[e, B]) {\cal Z}[e^i_k, B_k] = \int D
\Phi e^{-R[\Phi]} {\cal Z}[e^i_k[\Phi],B_k[\Phi]]\label{ZPSI12}
\end{equation}
with real - valued $R$. }
\end{remark}

\begin{remark}
Unlike the original Ho\v{r}ava construction of section \ref{sectHorava} the
action of Eq. (\ref{Se}) is written in coordinates $\bf Z$ that differ from
the original coordinates $\bf x$ of Eq. (\ref{FIn}). The fermions in the two
coordinates are related by an operator $e^{E_0 \, \hat \beta \,
t }\tilde{\Omega} \Pi \Omega^T$:
\begin{equation}
\Psi_{\bf Z} = e^{E_0 \, \hat \beta \, t }\tilde{\Omega} \Pi \Omega^T
\psi_{\bf x}
\end{equation}
Here  $\Omega$ brings operator $\hat A$ of Eq. (\ref{FIn}) to the block -
diagonal form of Eq. (\ref{blockdiag}). The corresponding coordinates are
denoted by $\cal P$ and may be identified with the coordinates of "momentum
space". $\Pi$ projects to the reduced four dimensional subspace of the $n$ -
component spinor space. Operator $\tilde{\Omega}$ commutes with $\hat \beta$
and relates spinors in "momentum space"  with spinors defined in the new
coordinates ${\bf Z}$. $E_0$ is the value of "energy" at the position of the
branches crossing.
\end{remark}

\begin{remark}
The considered above pattern of the emergent gravity and emergent $U(1)$
gauge field corresponds to the approximation, when the fields living
at various Fermi points are not correlated. The case, when such a
correlation appears complicates the pattern considerably. This may result in
the appearance of non - Abelian gauge fields \cite{VolovikBook} and the
generalization of the vierbein to the form, when the field $f^k_l$ becomes
matrix in flavor space. (Flavors enumerate Fermi points and the
corresponding Weyl spinors.) This remark is related also to the case of
section \ref{sectHorava}.

\revisionC{It is worth mentioning, that the action in the form of Eq.
(\ref{Se}) corresponds to the left - handed Weyl fermions in the presence of
the emergent vierbein $e^j_a$. For our purposes it is enough to consider
only the emergent left - handed fermions as the right handed ones are
related to them by charge conjugation. The situation, when the two emergent
left - handed fermions $\Upsilon_1, \Upsilon_2$ appear may be considered as
the appearance of one Dirac four - component spinor. Its left - handed
component is $\Upsilon_1$ while the right - handed component is defined as
$\epsilon_{AB} \bar{\Upsilon}_2^B$. When these two spinors are not
correlated we have two different vierbeins and two different $U(1)$ gauge
fields. If, for a certain reason, the two vierbeins coincide, then the two
different $U(1)$ gauge fields may be represented as one common vector $U(1)$
gauge field coupled to the Dirac fermion in a usual way and the second
common axial $U(1)$ gauge field that may alternatively be considered as an
axial component of torsion originated from spin connection.}

\end{remark}

\section{Conclusions}
\label{Discussion}

In this paper we discuss the
many-body
systems with multi - component
fermions. First of all, we consider in some details the particular case of
the  Ho\v{r}ava construction presented in \cite{Horava2005}, when the Fermi
surface of $3+1$ D model is reduced to the Fermi point. We prove theorem
\ref{theoremH}. It contains the original statement of Ho\v{r}ava given in
\cite{Horava2005} without proof. Namely, in the vicinity of the
topologically protected Fermi - point with topological invariant $N = 1$ the
emergent two - component Weyl spinors appear. In the case, when the fields
living in the vicinities of different Fermi points do not correlate with
each other, we may consider each low energy Weyl spinor separately. Then,
the emergent gravity given by the emergent vierbein appears that is
experienced by the Weyl fermions as the geometry of space, in which the
fermionic quasi - particles propagate. Besides, the emergent $U(1)$ gauge
field appears.

If the fields that belong to the vicinities of different Fermi points
correlate with each other, instead of the $U(1)$ gauge field the non -
Abelian gauge field may appear \cite{VolovikBook}. Besides, in this case the
vierbein is to be replaced by matrix in flavor space (flavor index
enumerates Fermi points and emergent Weyl spinors). The consideration of
this complication is out of the scope of the present paper.
The particular
problem, which requires further investigation, is:
what  global discrete or continuous symmetry of the underlying microscopic
theory (including the flavor symmetry) may reproduce the emergent gauge
symmetries of SM or GUT?

Then we consider the generalization of the problem discussed in section
\ref{sectHorava} to the case, when the original system contains multi -
component Majorana fermions. This case has been considered in section
\ref{sectMajorana}. The theorem \ref{theoremA} is proved, that is similar to
theorem \ref{theoremH}.
 Again, in the vicinity of the separate Fermi point the emergent two -
component Weyl spinor interacting with emergent vierbein and emergent U(1)
gauge field appears. The important difference from the case of section
\ref{sectHorava} is that the Weyl spinors emerge in space of generalized
coordinates $\bf Z$ that are different from the original coordinates $\bf
x$. \revisionC{Besides, in order to arrive at the model of emergent Weyl
fermions we suppose, that the
interactions do not break the fermion number conservation for the emergent
Weyl fermions. Remarkably, we do not need this requirement, when the
interaction between the fermions may be neglected. This suggests that there can be a special discrete symmetry in the underlying microscopic theory,
which forbids the violation of the fermion number conservation in the lowest order terms. The higher order terms may reflect the Majorana origin of the chiral Weyl particles, manifested in particular
in possibility of neutrinoless double beta decay}

The considered general constructions may have applications both in the
condensed matter physics and in the high energy physics. There  may exist
various condensed matter systems with multi - component fermions (both usual
ones and Majorana fermions) and with the Fermi - points. In particular,
certain Weyl semi - metals may belong to this class of systems. General
properties considered above predict, that the effective description of such
systems may be given in terms of the Weyl spinors interacting with emergent
gravity and emergent gauge field.

In the high energy theory the applications may be related to the paradigm,
in which Lorentz symmetry, the fermions that belong to its spinor
representations, the gravitational and gauge fields appear in the low energy
effective description of the underlined theory that works at extremely high
energies.
In the scenario, in which this theory contains multi - component Majorana
fermions, the observed coordinate space corresponds to the generalized
coordinates $\bf Z$, so that the coordinate space is the emergent
phenomenon, which follows from the matrix structure in momentum space (see
also \cite{CortesSmolin2013}).   Besides, the corresponding construction may
be related to the foundations of quantum mechanics.
The original Majorana fermion is described by the $n$ - component real -
valued wave function. The differential equation that describes its evolution
has the real - valued coefficients. The emergent low energy Weyl spinor, in
turn, is described by the complex - valued wave function. Thus, in this
pattern the complexification of quantum mechanics is the emergent low energy
phenomenon.

The work of M.A.Z. is  supported by the Natural Sciences and Engineering
Research Council of
Canada. GEV thanks Yu.G. Makhlin for fruitful discussions and
acknowledges the financial support  by the Academy of
Finland through its LTQ CoE grant (project $\#$250280).

\end{document}